\documentclass[aip,apl,reprint]{revtex4-1}
\usepackage{graphicx}
\usepackage{amsmath,amssymb}
\usepackage{bm}
\usepackage{epstopdf}

\begin{document}

\title{Perpendicular magnetic anisotropy in bulk and thin-film CuMnAs for antiferromagnetic memory applications}
\author{I. A. Zhuravlev}
\affiliation{Department of Physics and Astronomy and Nebraska Center for Materials and Nanoscience, University of Nebraska-Lincoln, Lincoln, Nebraska 68588, USA}

\author{A. Adhikari}
\affiliation{Department of Physics and Astronomy and Nebraska Center for Materials and Nanoscience, University of Nebraska-Lincoln, Lincoln, Nebraska 68588, USA}

\author{K. D. Belashchenko}
\affiliation{Department of Physics and Astronomy and Nebraska Center for Materials and Nanoscience, University of Nebraska-Lincoln, Lincoln, Nebraska 68588, USA}

\date{\today}

\begin{abstract}
CuMnAs with perpendicular magnetic anisotropy is proposed as an active material for antiferromagnetic memory. Information can be stored in the antiferromagnetic domain state, while writing and readout can rely on the existence of the surface magnetization. It is predicted, based on first-principles calculations, that easy-axis anisotropy can be achieved in bulk CuMnAs by substituting a few percent of As atoms by Ge, Si, Al, or B. This effect is attributed to the changing occupation of certain electronic bands near the Fermi level induced by the hole doping. The calculated temperature dependence of the magnetic anisotropy does not exhibit any anomalies. Thin CuMnAs(001) films are also predicted to have perpendicular magnetic anisotropy.
\end{abstract}

\maketitle

Antiferromagnets offer considerable benefits for storing information in spin-torque-based magnetic random-access memory (MRAM) cells compared to ferromagnets: the stray fields are minimized, and faster spin dynamics could be used to reduce the switching time. \cite{Wadley2016,Olejnik2018} Switching by current-induced spin-orbit torque and readout by anisotropic magnetoresistance have been demonstrated for CuMnAs \cite{Wadley2016} and Mn$_2$Au. \cite{Bodnar2018} In this scheme, information is stored as one of the two orthogonal orientations of the antiferromagnetic order parameter. The latter always lies in the plane of the film, perpendicular to the tetragonal axis, because the magnetocrystalline anisotropy (MCA) of both CuMnAs \cite{Wadley2013a} and Mn$_2$Au \cite{Shick2010} is easy-plane. Switching is achieved by passing an electric current along one of the two orthogonal directions. The current generates a field-like spin-orbit torque, aligning the order parameter $\mathbf{L}=\mathbf{m}_1-\mathbf{m}_2$ (where $\mathbf{m}_i$ are the sublattice magnetizations) perpendicular to the direction of the current flow. The simplest form of the current-induced staggered effective field in CuMnAs or Mn$_2$Au is derivable from the effective potential $U=\alpha \mathbf{L}\cdot(\mathbf{E}\times\mathbf{z})$. Terms of this kind are forbidden in all antiferromagnets whose magnetic space group contains an antitranslation, i.e., a time-reversal operation combined with a lattice translation.

Easy-plane MCA is an obstacle for scaling the memory density, because the two orthogonal in-plane orientations of the order parameter are separated by a very small energy barrier proportional to the fourth-order in-plane anisotropy. Thus, the memory bits are expected to lose thermal stability as their size is reduced toward the nanoscale. Perpendicular (easy-axis) anisotropy is favorable for thermal stability, but it does not allow the switching and readout scheme of Ref.\ \onlinecite{Wadley2016}. However, alternative switching and readout schemes are possible for CuMnAs and Mn$_2$Au with perpendicular anisotropy thanks to their special symmetry properties.

\begin{figure}[htb]
\includegraphics[width=0.9\columnwidth]{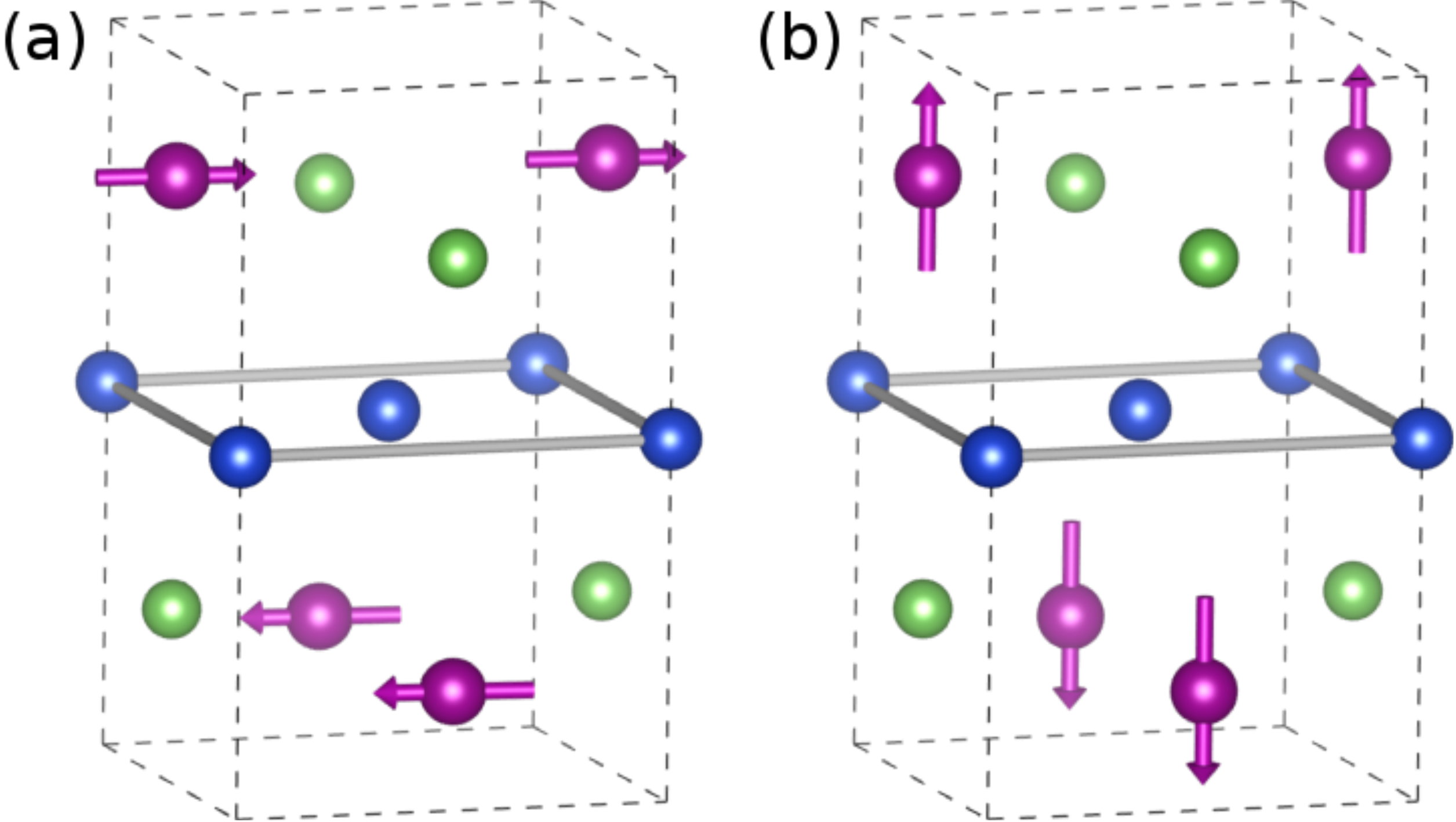}
\caption{Magnetic unit cell of tetragonal CuMnAs. Dashed lines show the cell boundary. Mn, As, and Cu atoms are shown by purple, green, and blue spheres, respectively. (a) Mn spins lie in the basal plane. (b) Mn spins are parallel to the tetragonal axis. The figure also represents the quintuple-layer building block used to model a CuMnAs(001) film.}
\label{fig:structure}
\end{figure}

Tetragonal CuMnAs or Mn$_2$Au with easy-axis anisotropy would have two antiferromagnetic domains related to each other by the time-reversal operation. Because the magnetic structure \cite{Wadley2015} of these materials (see Fig.\ \ref{fig:structure} for CuMnAs) has no antitranslation, these domain states are macroscopically distinguishable and can be used as a robust binary state variable to store information. This feature is shared with insulating magnetoelectric antiferromagnets like Cr$_2$O$_3$. \cite{He2010,Andreev,Belashchenko2010} Furthermore, symmetry requires that single-domain CuMnAs or Mn$_2$Au must have an uncompensated surface magnetization strongly coupled to the antiferromagnetic order parameter, even if the surface is not atomically flat. \cite{Belashchenko2010} In an AF/N/AF device, where AF is an antiferromagnet and N a normal metal or tunnel barrier, the presence of a surface magnetization makes the antiferromagnet act essentially as a ferromagnet. In particular, a first-principles calculation for a CuMnAs/GaP/CuMnAs tunnel junction predicted both sizeable tunneling magnetoresistance (TMR) and spin-transfer torque (STT) in a noncollinear spin configuration. \cite{Stamenova} As in a ferromagnetic STT-MRAM cell, the current-induced torque can be used to switch the domain state of one of the antiferromagnetic layers, while TMR can be used for readout.

Having this target device in mind, we now predict, using first-principles calculations, that the magnetic anisotropy of CuMnAs can be turned from easy-plane [Fig.\ \ref{fig:structure}(a)] to easy-axis [Fig.\ \ref{fig:structure}(b)] by a small substitutional doping on the As sublattice or by using a very thin film.

\begin{figure}[htb]
\includegraphics[width=0.9\columnwidth]{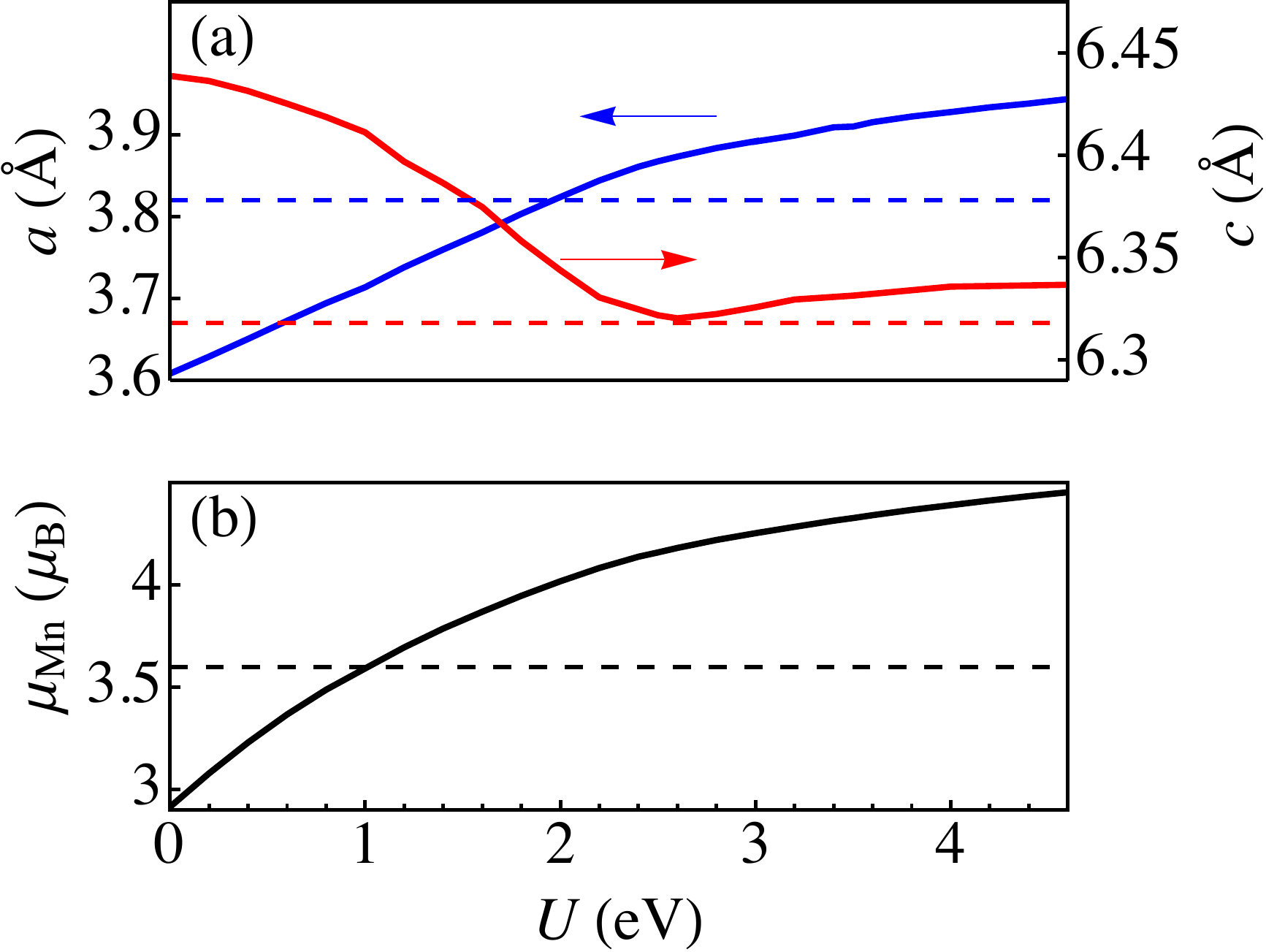}
\caption{(a) Equilibrium lattice parameters $a$ and $c$ and (b) local moments $\mu_{\mathrm{Mn}}$ on the Mn atoms in CuMnAs as a function of $U$, with $J=0.58$ eV. Dashed lines: experimental data at room temperature.
\cite{Wadley2013,Wadley2013a}}
\label{fig:vaspacmom}
\end{figure}

Using the generalized gradient approximation (GGA)\cite{PBE} results in a considerable disagreement with experiment in the lattice constants and sublattice magnetization. Therefore, as a first step, we use the GGA$+U$ method in the Vienna ab-initio simulation package (VASP) \cite{VASP} and empirically adjust the value of $U$ while keeping $J$ fixed at 0.58 eV. The equilibrium lattice constants and the local magnetic moment on the Mn atoms are shown in Fig.\ \ref{fig:vaspacmom} as a function of $U$. Good agreement with experiment is achieved for $U=2.5$ eV. The value $U-J\approx 2$ eV also gives good agreement with photoemission measurements. \cite{Veis2018} Therefore, we adopt $U=2.5$ eV for band structure calculations. The corresponding structural parameters are $a=3.872$ \AA, $c=6.326$ \AA, $ z_\mathrm{Mn}$=0.6614, and $z_\mathrm{As}=0.2615$.

\begin{figure}[htb]
\includegraphics[width=0.9\columnwidth]{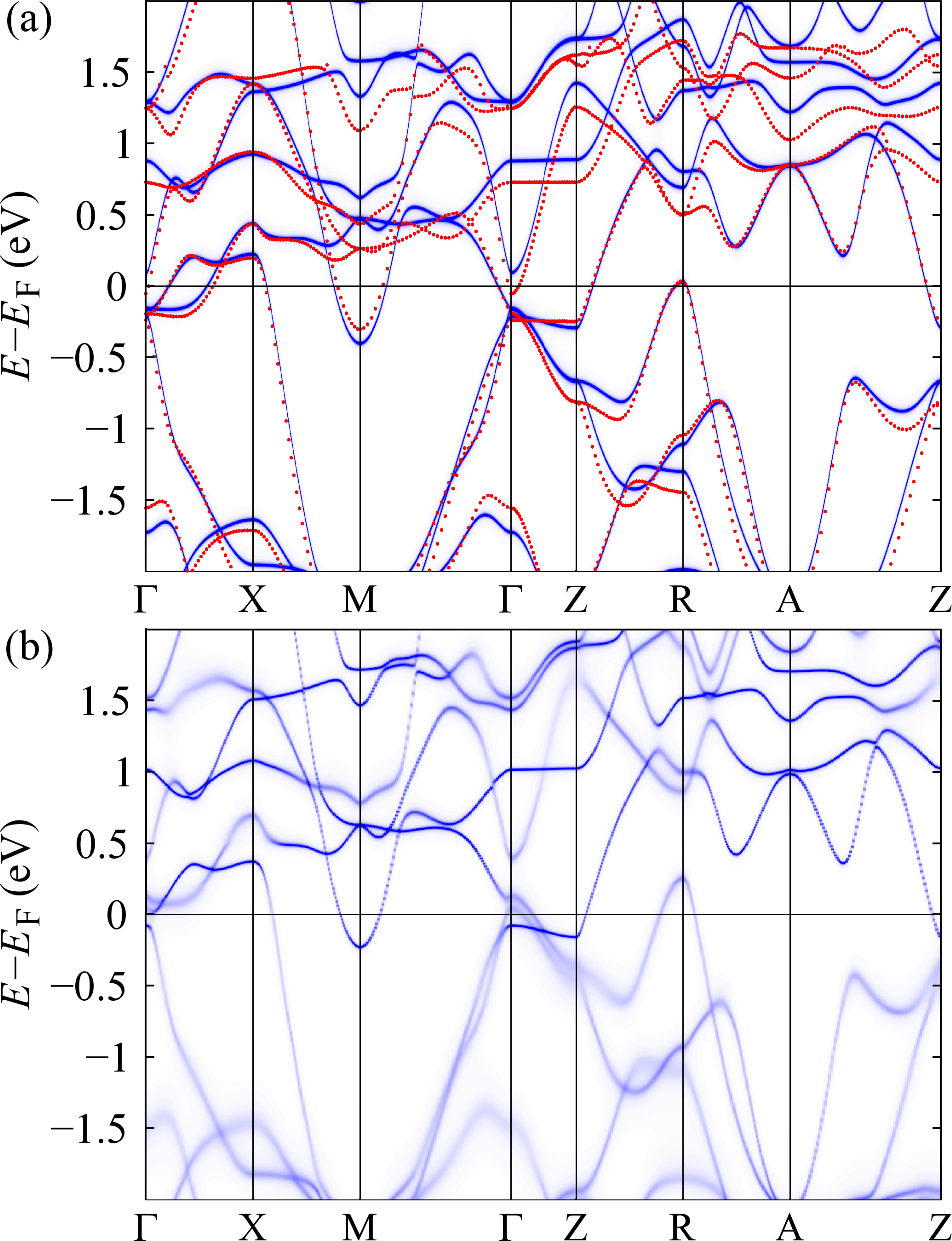}
\caption{(a) CuMnAs band structure calculated using VASP (red dots) with $U=2.5$ eV and GF-LMTO (blue lines) with GGA and scaled exchange-correlation field (see text).
(b) Spectral function of the CuMnAs$_{0.9}$Si$_{0.1}$ alloy (CPA calculation with spin-orbit coupling).}
\label{fig:bndvasplmgf}
\end{figure}

To study substitutional alloys, we use the coherent potential approximation (CPA) implemented within the Green's function-based tight-binding linear muffin-tin orbital (GF-LMTO) method. \cite{Turek,CPA} The main effect of the on-site Coulomb correlations is to increase the spin splitting in the half-filled $3d$ shell of the Mn atoms. In the GF-LMTO calculations, this effect is mimicked by using GGA \cite{PBE} (instead of GGA$+U$) with the local part of the exchange-correlation field for Mn atoms scaled by a factor 1.28. This factor, as well as the radii of the atomic spheres, have been adjusted \cite{ASA-note} to make the LMTO band structure reproduce the VASP results as closely as possible; this comparison is illustrated by Fig. \ref{fig:bndvasplmgf}(a).

The spin-orbit coupling was included as a perturbation of the LMTO potential parameters, \cite{Turek2008,Belashchenko2015,Zhuravlev2015} and the MCA energy $K$ was obtained by calculating the single-particle energy difference for the in-plane and out-of-plane orientations of the antiferromagnetic order parameter with the charge density taken from the self-consistent calculation without spin-orbit coupling. The reciprocal-space integration was converged for the uniform $30\times30\times18$ mesh in the full Brillouin zone.

\begin{figure}[htb]
\includegraphics[width=0.85\columnwidth]{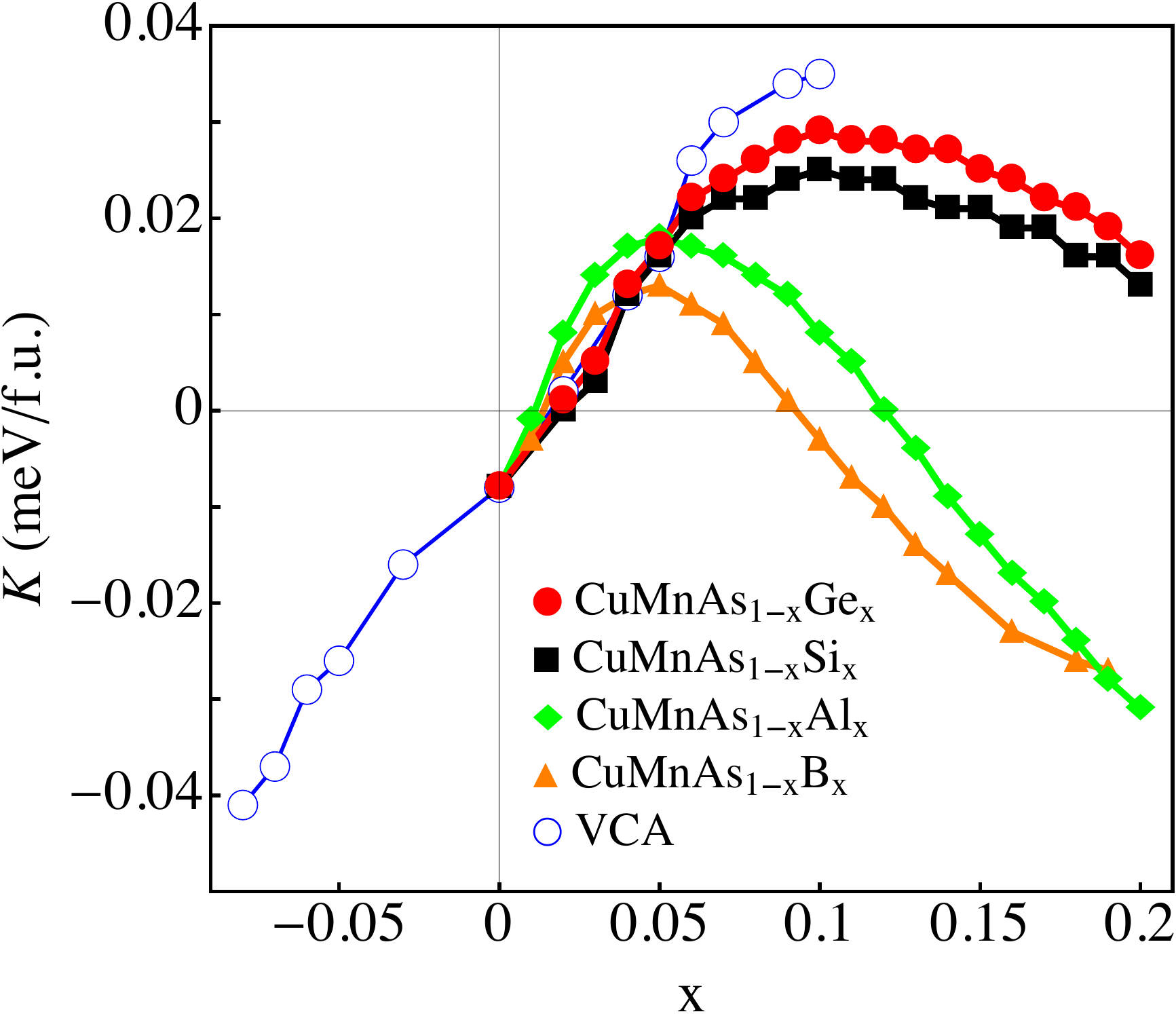}
\caption{Concentration dependence of MCA in CuMnAs$_{1-x}$Y$_{x}$ with $\mathrm{Y} = \mathrm{Ge}$, Si, Al, or B, calculated in CPA. Empty blue circles: MCA in VCA with $x$ representing the number of holes per As atom.}
\label{fig:MAEConc}
\end{figure}

We focus on the substitution of $sp$ elements on the As sublattice in order to avoid strong band broadening, which tends to suppress MCA.\cite{Turek-MCA,Belashchenko2015,Zhuravlev2015} Figure \ref{fig:MAEConc} shows the concentration dependence of MCA calculated using GF-LMTO and CPA for CuMnAs$_{1-x}$Y$_{x}$, where Y stands for Ge, Si, Al, or B. Small concentrations of any of these dopants are found to induce a spin-reorientation transition to easy-axis anisotropy. The MCA increases with increasing concentration of the dopants up to about 10\% for Ge and Si or 5\% for Al and B. The largest easy-axis MCA is obtained with 10\% Ge or Si substitution for As.

The fact that maximum MCA is reached at a twice smaller concentration of Al or B compared to Ge or Si strongly suggests that the changes in MCA are primarily due to the effect of alloying on band filling. Indeed, Al and B donate twice as many holes as Ge and Si. This conclusion is supported by the fact that the MCA calculated in the virtual crystal approximation (VCA) closely follows the CPA results for the same hole doping level. The reason for this similarity becomes clear on inspection of the spectral function for CuMnAs$_{0.9}$Si$_{0.1}$, which is shown in Fig. \ref{fig:bndvasplmgf}(b). The bands are not strongly broadened, and the main difference with the band structure of pure CuMnAs is in the position of the Fermi level.

\begin{figure}[htb]
\includegraphics[width=0.9\columnwidth]{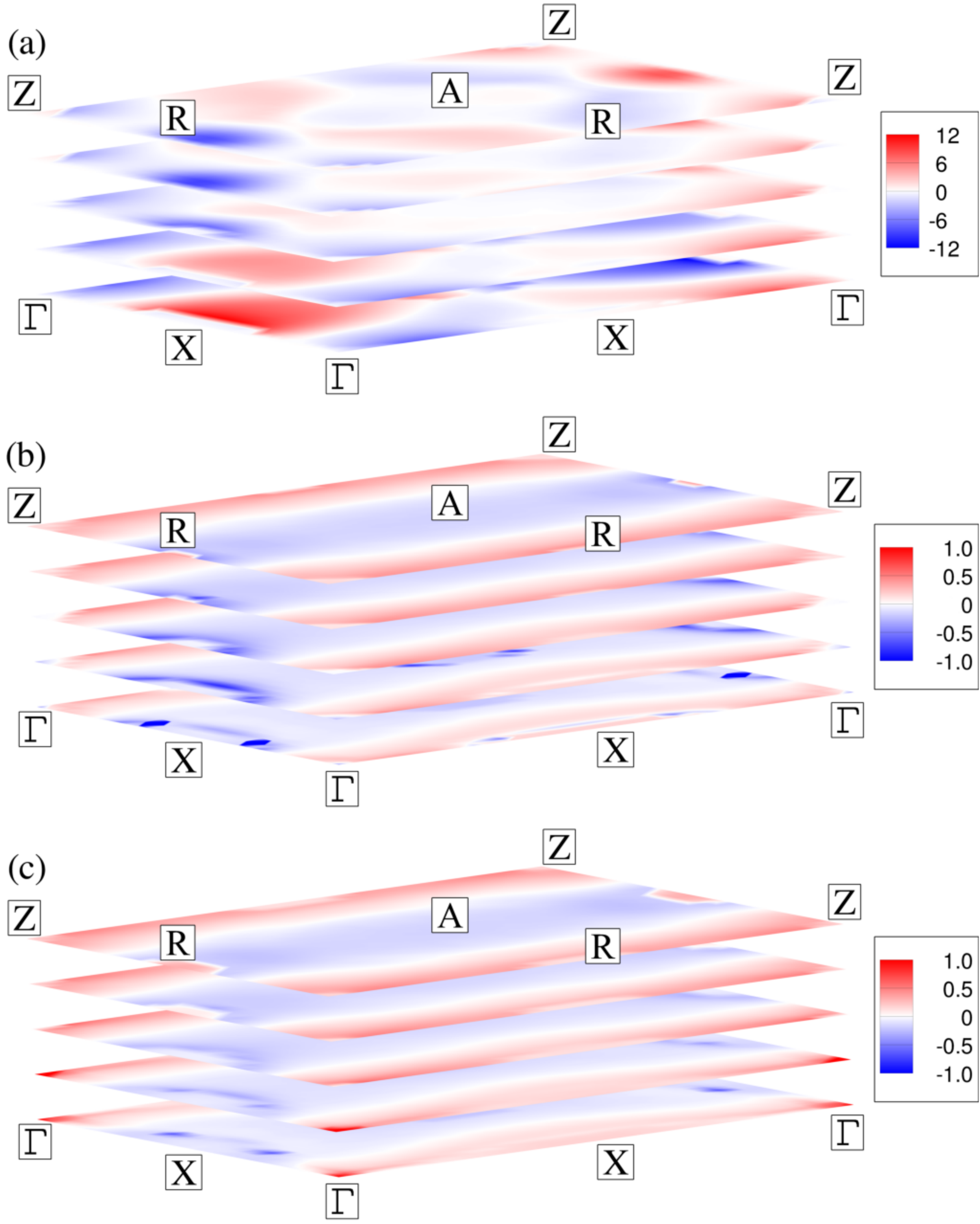}
\caption{Wave vector-resolved MCA (eV$\cdot a_0^3$) in CuMnAs$_{1-x}$Si$_{x}$. (a) $x=0$, without symmetrization; (b) $x=0$, with symmetrization; (c) $x=0.1$, with symmetrization.}
\label{fig:MAEKres}
\end{figure}

For further insight, it is useful to analyze the contributions to MCA resolved in reciprocal space. \cite{Belashchenko2015,Zhuravlev2015} Figure \ref{fig:MAEKres}(a) shows such $\mathbf{k}$-resolved MCA in CuMnAs, defined as the difference between the single-particle energies $E_{sp}(\mathbf{k})=\sum_{i, \epsilon_{i\mathbf{k}}<\epsilon_F} (\epsilon_{i\mathbf{k}}-\epsilon_F)$ for $\mathbf{L}\parallel\mathbf{y}$ and $\mathbf{L}\parallel\mathbf{z}$. The interpretation of this figure is, however, complicated by the presence of large contributions of opposite sign that are approximately odd in $k_x$. The reasons is that, in the absence of both inversion and time-reversal symmetry, the spin-orbit-coupled band structure is not generally even with respect to $\mathbf{k}$. In particular, for $\mathbf{L}\parallel\mathbf{y}$ there is no magnetic symmetry operation that reverses the sign of $k_x$ (or of both $k_x$ and $k_y$). It is, therefore, useful to symmetrize the $\mathbf{k}$-resolved anisotropy with respect to $k_x$. The result is shown in Fig.\ \ref{fig:MAEKres}(b); note the greatly reduced scale compared to Fig.\ \ref{fig:MAEKres}(a).

Fig.\ \ref{fig:MAEKres}(b) reveals large negative contributions to MCA coming from distinct ``hot spots'' \cite{Zhuravlev2015,Kondorskii} in the vicinity of the $\Gamma$X line. This negative contribution can be attributed to the electronic bands crossing the Fermi level along the $\Gamma$X line [see Fig.\ \ref{fig:bndvasplmgf}(a)]. Hole doping suppresses the negative hot spots and instead gives rise to large positive contributions near $\Gamma$, as can be seen from Fig.\ \ref{fig:MAEKres}(c) for CuMnAs$_{0.9}$Si$_{0.1}$. Comparing the spectral functions for pure and Si-doped CuMnAs in Fig.\ \ref{fig:bndvasplmgf}(a-b), we see that the relevant bands are shifted above the Fermi level under hole doping. The positive contribution near $\Gamma$ appears thanks to the presence of several closely spaced bands around the Fermi level. Thus, the upward shift of the bands relative to the Fermi level leads to the spin-reorientation transition under small hole doping.

\begin{figure}[htb]
\includegraphics[width=0.9\columnwidth]{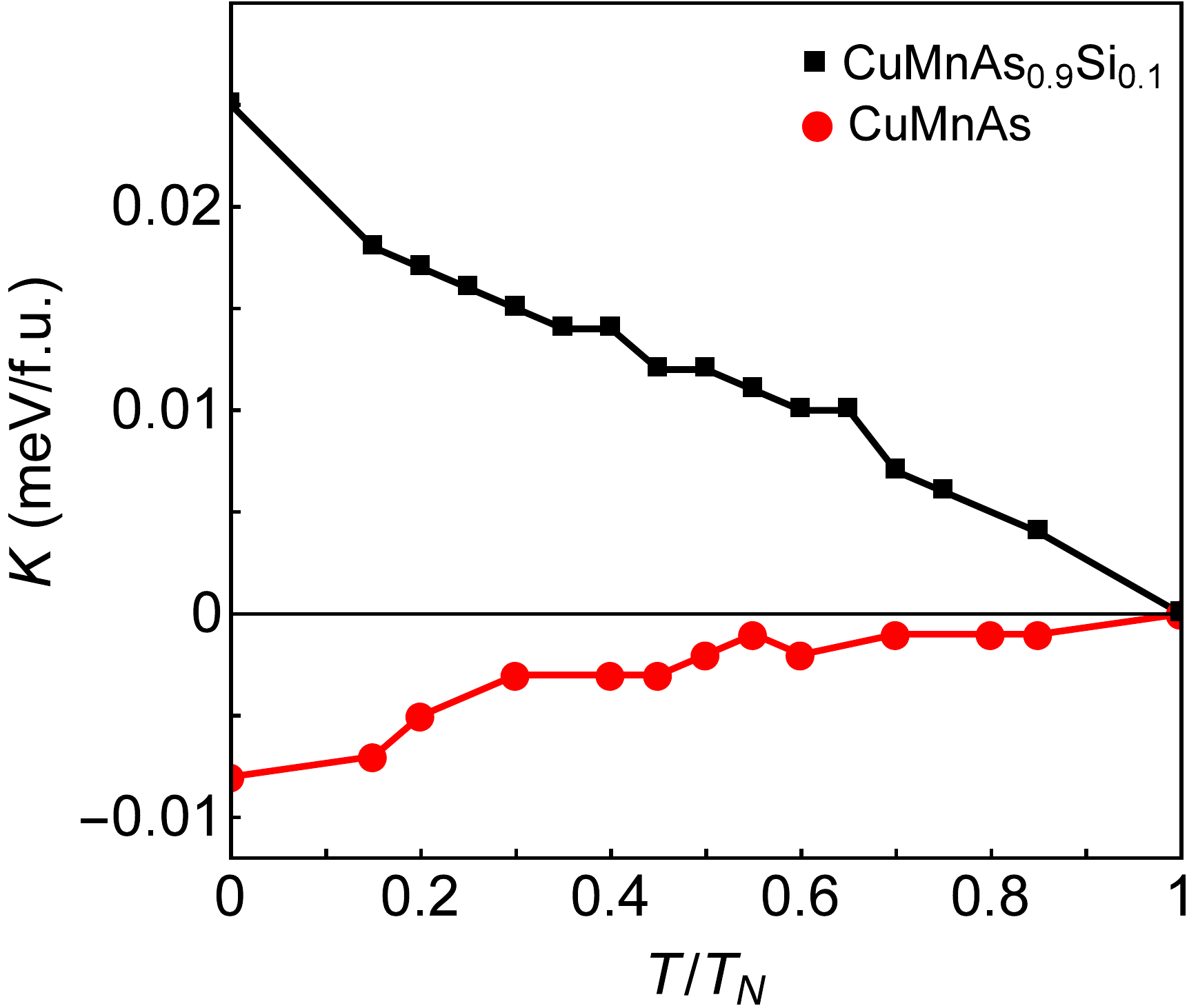}
\caption{MCA in CuMnAs (red circles) and CuMnAs$_{0.9}$Si$_{0.1}$ (black squares) as a function of the reduced temperature $T/T_N$.}
\label{fig:MAETemp}
\end{figure}

For device applications, it is important to know whether the results described above remain valid at room temperature. Since CuMnAs is an itinerant metal, the temperature dependence of its MCA could be non-trivial. \cite{Zhuravlev2015,Chang2018} To check for possible anomalies, we have calculated the dependence of MCA in CuMnAs and CuMnAs$_{0.9}$Si$_{0.1}$ on the reduced temperature $T/T_N$ (where $T_N$ is the N\'eel temperature) using the disordered local moment (DLM) method \cite{DLM1,DLM2} implemented as described in Ref.\ \onlinecite{Zhuravlev2015}. Fig.\ \ref{fig:MAETemp} shows that in both cases the temperature dependence is monotonic. We have checked that the variation of the unit cell volume and the $c/a$ ratio that may be expected from thermal expansion have only a weak effect on MCA.

\begin{figure}[htb]
\includegraphics[width=0.9\columnwidth]{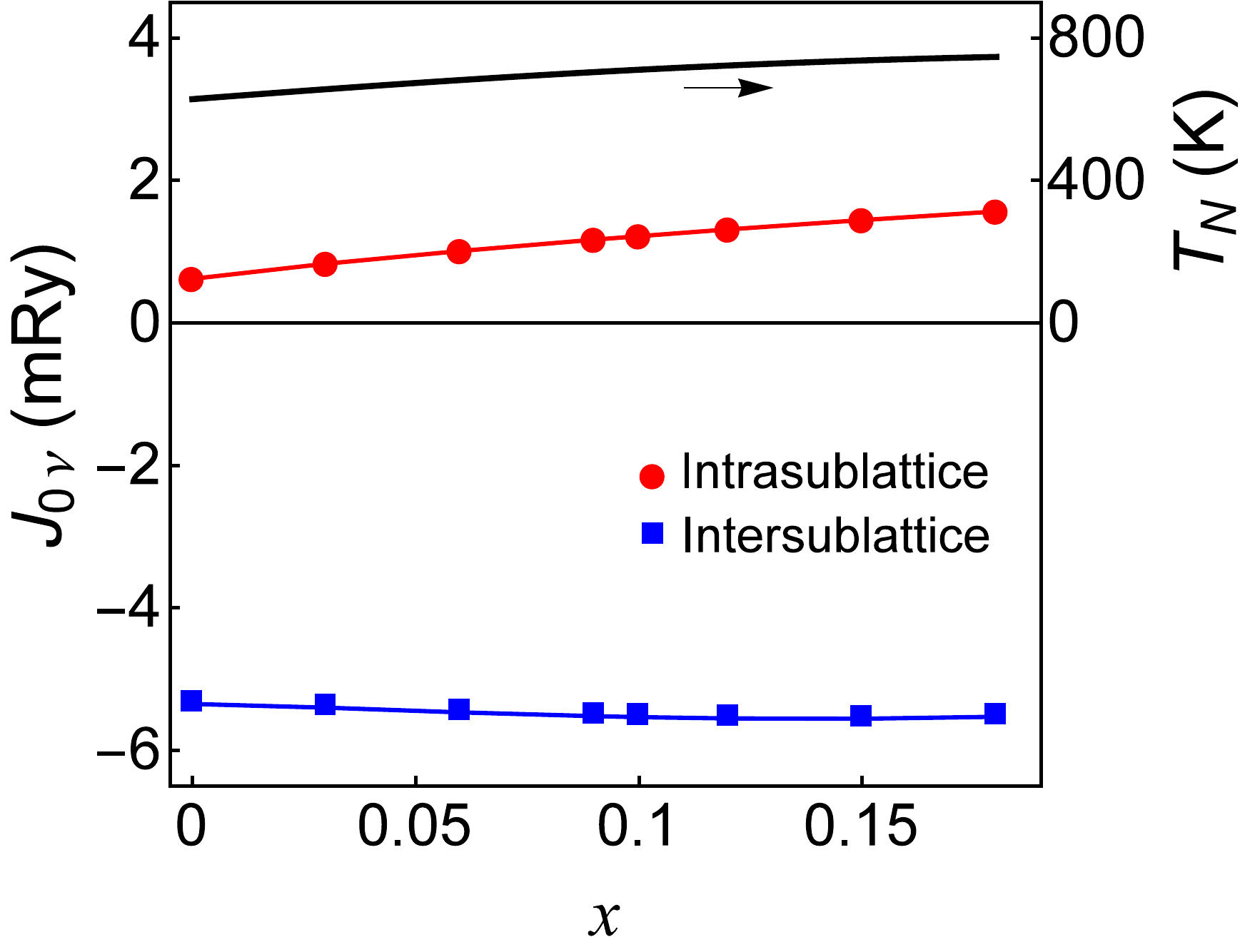}
\caption{Calculated exchange parameters $J_{0\nu}$ (symbols) and mean-field $T_{N}$ (solid black line) in CuMnAs$_{1-x}$Si$_{x}$. Red circles and blue squares: total intra- and intersublattice exchange $J_{0\nu}$, respectively.}
\label{fig:JijTn}
\end{figure}

Further, Fig.\ \ref{fig:JijTn} shows the concentration dependence of the exchange parameters in CuMnAs$_{1-x}$Si$_{x}$ calculated using the linear response method in the paramagnetic state. \cite{exchange,DLM2,Turek-exch} The two curves show the sums $J_{0\nu}=\sum_{j\in\nu} J_{ij}$ of the exchange parameters $J_{ij}$, where one Mn site $i$ is fixed while the other $j$ is restricted to belong either to the same sublattice as $i$ or to the other sublattice. The exchange coupling parameters at $x=0$ are in reasonable agreement with earlier calculations, \cite{Maca2017} and correspond to the correct antiferromagnetic ground state of CuMnAs. It is seen that the strong intersublattice exchange coupling remains almost constant under Si doping, while the weaker intrasublattice coupling strengthens. The N\'eel temperature $T_N$ calculated in the mean-field approximation shows a moderate increase with Si doping. In pure CuMnAs it is overestimated by 31\% compared to the experimental value of 480 K,\cite{Hills} as expected of this approximation.

Taken together, the monotonic temperature dependence of MCA and the weak dependence of the dominant exchange coupling and $T_N$ on the concentration suggest that the conclusions about the concentration dependence of MCA drawn from Fig.\ \ref{fig:MAEConc} remain valid at room temperature.

Table \ref{dH} lists the formation enthalpies of the substitutional impurities, calculated using elemental solids as a reference (we used the $\alpha$-rhombohedral phase for B). It is seen that the formation enthalpies for Si and Ge are quite small, and these elements should readily substitute for As. The formation enthalpies for Al and B are also not prohibitively high.
\begin{table}[htb]
\caption{Formation enthalpies $\Delta H_f$ (eV) of substitutional impurities in CuMnAs.}
\begin{tabular}{|l|c|c|c|c|}
\hline
Atom & B & Al & Si & Ge \\
\hline
$\Delta H_f$ & 1.16 & 0.55 & 0.03 & 0.12 \\
\hline
\end{tabular}
\label{dH}
\end{table}

The thickness of magnetic layers in MRAM devices is typically a few nanometers, and the magnetic anisotropy of a thin film can be dominated by surface or interface anisotropy. \cite{Carcia} Therefore, we now consider the magnetic anisotropy of a thin CuMnAs(001) film. It is clear from Fig.\ \ref{fig:structure} that the tetragonal structure of CuMnAs can be viewed as a stacking of stoichiometric blocks consisting of two buckled MnAs layers sandwiching a Cu layer. Below we refer to these blocks as quintuple layers. We model a thin CuMnAs film as a stack of several quintuple layers, which maintains the overall stoichiometry of the film and makes its two surfaces equivalent. The films are terminated on each side by a layer of magnetic Mn atoms.

\begin{figure}[htb]
\includegraphics[width=0.85\columnwidth]{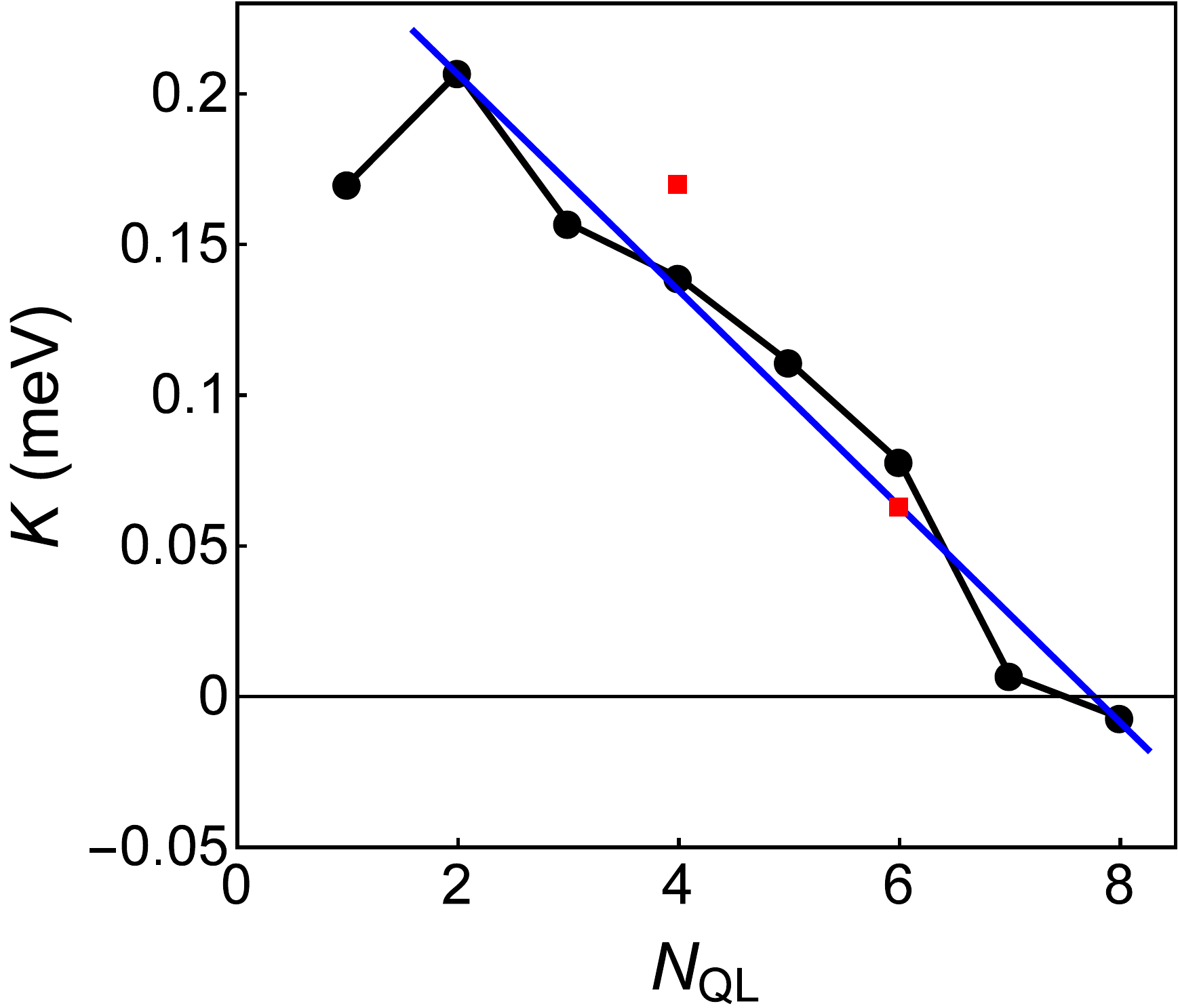}
\caption{Total MCA of a CuMnAs slab as a function of its thickness measured in the number of quintuple layers $N_\mathrm{QL}$. Black circles: with unrelaxed atomic positions from the bulk structure. Red squares: with the relaxed structure constrained to have the bulk lattice parameter in the basal plane. Blue line: linear fit to the unrelaxed data for $N_\mathrm{QL}\geq2$.}
\label{fig:MAEslab}
\end{figure}

Fig.\ \ref{fig:MAEslab} shows the MCA of a CuMnAs film as a function of its thickness, measured in the number of quintuple layers, calculated in VASP. The vacuum layer used in the periodic setup had a thickness of 8.4 \AA.

The results shown by black circles in Fig.\ \ref{fig:MAEslab} were obtained using the unrelaxed atomic positions obtained by terminating the bulk structure. For $N_\mathrm{QL}=4$ and 6 we also performed a full relaxation while keeping the lattice parameter in the basal plane fixed; the corresponding results are shown by red squares in Fig.\ \ref{fig:MAEslab}. It is seen that structural relaxation at the surface has a rather small effect on MCA.

Fig.\ \ref{fig:MAEslab} shows that MCA is positive in CuMnAs films whose thickness is less than about 7 quintuple layers, or about 4 nm. The thickness dependence is approximately linear, with the vertical intercept giving the effective surface anisotropy and the slope corresponding to bulk MCA. The fitted slope is about $-0.018$ meV/f.u., which is in good agreement with $-0.022$ meV/f.u. obtained from the bulk calculation. (This value is of the same sign but of greater magnitude compared to the GF-LMTO result.) Overall, these results suggest that very thin CuMnAs(001) films have perpendicular magnetic anisotropy, which can be exploited in spintronic devices.

In conclusion, we argue that CuMnAs with perpendicular magnetic anisotropy can be used as an active element in an antiferromagnetic memory cell, storing information in its domain state, with the surface magnetization facilitating spin-transfer-torque switching and magnetoresistive readout. We predict that easy-axis anisotropy in bulk CuMnAs can be achieved by substituting a few percent of Ge, Si, Al, or B for As. We also predict that thin CuMnAs films up to a few nanometers thick also have perpendicular anisotropy.

This work was supported by the National Science Foundation through Grant No. DMR-1609776 and the Nebraska MRSEC (Grant No. DMR-1420645) and performed utilizing the Holland Computing Center of the University of Nebraska.


\begin{thebibliography}{99}
\bibitem{ASA-note} Sphere radii of 3.02, 2.95, and 2.72 a.u.\ and basis sets with $l\leq 3$ were used for Mn, Cu, and As, respectively. In addition, empty spheres with a radius of 1.45 a.u.\ and $l\leq2$ were placed at the corners and base centers of the unit cell shown in Fig.\ \ref{fig:structure}.
\end{thebibliography}
\end{document}